\documentclass[sigconf,nonacm]{acmart}

\AtBeginDocument{%
  \providecommand\BibTeX{{\normalfont B\kern-0.5em{\scshape i\kern-0.25em b}\kern-0.8em\TeX}}}

\setcopyright{none}
\acmConference[Preprint]{Preprint}{2026}{}
\renewcommand\footnotetextcopyrightpermission[1]{}

\usepackage{booktabs}
\usepackage{graphicx}

\begin{document}

\title{SilentProbe: Measuring Silent Failure in Production APIs
       Used as Agent Tools}

\author{Zongrong Li}
\affiliation{%
  \institution{Department of Geography, Texas A\&M University}
  \city{College Station}
  \state{Texas}
  \country{USA}
}
\affiliation{%
  \institution{Monid, Inc.}
  \city{San Francisco}
  \state{California}
  \country{USA}
}
\email{zongrong@tamu.edu}

\author{Shengkun Ye}
\authornote{Corresponding author.}
\affiliation{%
  \institution{Monid, Inc.}
  \city{San Francisco}
  \state{California}
  \country{USA}
}
\email{shengkun@monid.ai}

\author{Feiyou Guo}
\authornotemark[1]
\affiliation{%
  \institution{Monid, Inc.}
  \city{San Francisco}
  \state{California}
  \country{USA}
}
\email{feiyoug@monid.ai}

\author{Zuoyou Dang}
\affiliation{%
  \institution{Monid, Inc.}
  \city{San Francisco}
  \state{California}
  \country{USA}
}
\email{ooctoo@monid.ai}

\begin{abstract}
An LLM agent calling a production API cannot distinguish a query that matched
nothing from a query the server did not understand. Both return HTTP 200 with a
parsable body, no exception to catch and no field to branch on. We ask what
predicts which one occurred, and what it does to the agent.

Auditing 721{,}320 parameters across 2{,}501 independently published OpenAPI
documents, we find that 7.5\% declare an enumeration and 15.2\% declare any
machine-checkable constraint at all, while 40.1\% of documents state at least
one constraint in prose that their schema does not encode. Executing 219
schema-derived perturbations against live commercial endpoints from 27 vendors,
reached through a single aggregation layer (Monid) that publishes a schema and
returns a run identifier for every call, we find that constraint form, not vendor
identity, predicts honesty:
machine-checkable constraints yielded an honest error in 111 of 111 cases,
prose-only constraints failed silently in 44 of 61 ($p=2\times10^{-13}$).

Twelve models across eight families then met these endpoints on ordinary tasks.
A vocabulary that the description merely exemplifies was missed by every model
on 88 of 88 attempts, while vocabularies written out in full were used correctly
88 to 91\% of the time. Running the full agent loop, models detected the
resulting silent failure in 12\% of cases, repaired it in 0\%, asserted a false
negative to the user in 41\%, and invented a figure in 12\%. Promoting the
vocabulary into the schema removes the failure, from 88 of 88 to 0 of 89. The
fix is one line of schema rather than a better model.

Code, schemas, perturbation sets, agent transcripts and per-call run identifiers
are released at \url{https://github.com/Jasper0122/silentprobe}.
\end{abstract}



\maketitle

\section{Introduction}

An LLM agent issuing a query against a third-party API cannot tell these two
situations apart:

\begin{enumerate}
\item the query was well formed and no records match it, and
\item the query was not understood, the server discarded part of it, and the
      response describes a different question.
\end{enumerate}

\noindent
Both arrive as HTTP 200 with a body the agent can parse. There is no exception to
catch, no status code to branch on, and no field in the response that marks the
difference. The model has done nothing wrong. The environment has misled it by
omission.

This is not hallucination and it is not a model capability problem. It is a
property of the interface contract. A parameter's accepted values are frequently
described in a sentence meant for a human reader rather than declared in the
machine-readable schema, and a sentence is something only the caller can act on.
No validator, no gateway and no framework can enforce it. When a model emits a
reasonable term the vendor happens not to use, the request is syntactically
valid, passes every automated check, reaches the vendor, and comes back empty.

\subsection{A motivating observation}
\label{sec:motivating}

Table~\ref{tab:seed} shows four calls to one commercial people-search endpoint,
differing only in the spelling and format of two parameters. All four returned
HTTP 200 with no error field.

\begin{table}[t]
\caption{Four calls to one endpoint, varying only parameter spelling and format.
         All returned HTTP 200 with no error field. The vendor's controlled
         vocabulary contains \texttt{vp} and its range format uses a comma.}
\label{tab:seed}
\begin{tabular}{clll}
\toprule
\# & \texttt{seniority} & \texttt{employees\_range} & Results \\
\midrule
1 & \texttt{vp}              & \texttt{"51,200"} & 28{,}417 \\
2 & \texttt{vice\_president} & \texttt{"51,200"} & \textbf{0} \\
3 & \texttt{vp}              & \texttt{"51-200"} & 160{,}884 \\
4 & \texttt{vp}              & \emph{omitted}    & 160{,}884 \\
\bottomrule
\end{tabular}
\end{table}

Two distinct failures appear here. \textbf{Row 2: a synonym outside the
vocabulary returns zero, not an error.} \texttt{vice\_president} is a reasonable
guess for any caller that has not memorised the vendor's terminology, and the
published schema does not forbid it because the schema declares the field only as
a string. The result reads as ``there are no vice presidents at companies of this
size.''

\textbf{Rows 3 and 4: a malformed range is discarded, not rejected.} These return
the identical count, and that equality is the load-bearing evidence: a filter
contributing nothing returns exactly what no filter returns. The server accepted
a constraint, ignored it, and reported success. Row 3's 160{,}884 records are not
a harmless superset; most are companies whose headcount falls outside the range
the caller asked for.

The second failure is the more dangerous. An empty result at least looks
suspicious to a careful consumer. A large, plausible, well-formed result that
silently violates a stated constraint does not.

\subsection{Approach and contributions}

The obvious question is how often production APIs behave this way. That question
produces an answer which is largely a property of whichever endpoints one happens
to sample. The more useful question, and the one this paper answers, is
\emph{what predicts it}. The answer is readable from published artefacts before a
single call is made, and it is fixable.

There is also a practical reason this measurement has not been made before.
Comparing failure behaviour across vendors means executing thousands of
deliberately malformed calls against commercial APIs, under one contract, with
schemas in a common format, and with a per-call record of what each request
returned. Assembled the usual way that is separate commercial agreements,
separate authentication, incompatible specification formats, and no shared
identifier by which a single measurement can later be cited. We avoid it by
working through an aggregation layer, Monid~\cite{monidapi}, which supplies
those four properties at once; Section~\ref{sec:method} states which of them
the method depends on, and Section~\ref{sec:discussion} treats the resulting
constraint on the sample as a limitation.

\begin{enumerate}
\item A taxonomy of silent failure modes for production APIs used as agent tools,
      including two that the agent-robustness literature's noise models omit: the
      \emph{silent drop}, where a discarded filter returns a plausible superset,
      and the \emph{semantic downgrade}, where the caller weakens its own request
      to whatever the documented vocabulary can express.
\item A free, static audit detecting constraints stated in prose but absent from
      the schema, run over two independent corpora: a commercial aggregation
      layer and 2{,}501 public OpenAPI documents.
\item Evidence from 219 executed perturbations across 27 vendors that
      \emph{constraint form}, not vendor identity, predicts whether a malformed
      call fails honestly or silently, and that actionable error messages are
      themselves a product of machine-readable schemas.
\item An end-to-end study over twelve models showing that disclosure and
      machine-readability are separable: models comply with vocabularies written
      out in prose and fail universally on vocabularies shown only by example.
\item Measurement of downstream harm in full agent loops, and an intervention
      with a measured effect together with the boundary conditions under which it
      is unnecessary.
\end{enumerate}

All code, schemas, perturbation sets, agent transcripts and judge labels are
released at \url{https://github.com/Jasper0122/silentprobe}. Every executed call carries a run identifier, and all are
included, so any individual measurement reported here can be re-fetched rather
than taken on trust.

\section{Related Work}
\label{sec:related}

\paragraph{Agent tool-use benchmarks assume an honest environment}
ToolBench aggregates 3{,}451 tools and 16{,}464 REST APIs from
RapidAPI~\cite{qin2024toolllm}; API-Bank provides 73 executable APIs
reimplemented in Python~\cite{li2023apibank}; MCP-Bench connects agents to 28
live MCP servers spanning 250 tools~\cite{mcpbench2025}; BFCL evaluates function
calling across single-turn, multi-turn and multi-step
categories~\cite{patil2025bfcl}. These differ substantially in how real their
environments are. What they share is the scoring target: whether the trajectory
reached the right answer. Whether the environment answered the question that was
asked is never measured. API-Bank's locally reimplemented endpoints cannot
exhibit vendor-specific parsing behaviour at all, because there is no vendor.

\paragraph{The response to unreliable environments was to simulate them}
ToolBench's APIs are executed for real, and are unstable in consequence. That
instability motivated StableToolBench, which adds a caching layer plus a
GPT-4-turbo API simulator~\cite{guo2024stabletoolbench}, and MirrorAPI, which
trains an environment model over request/response pairs from more than 7{,}000
RapidAPI endpoints~\cite{mirrorapi2025}. Both are careful engineering, and both
are fatal to the measurement we want to make. A simulator produces responses
consistent with an endpoint's documentation, and no documentation states that a
parameter will be silently discarded. A benchmark built on simulated tools
reports a robustness it has not tested.

\paragraph{Agent robustness work injects a distribution it has not measured}
AgentNoiseBench generates tool-side noise using a frozen model as an adversarial
noise generator, over simulated environments, and reports no real-world
rate~\cite{agentnoisebench2026}; related work applies chaos-engineering-style
fault injection~\cite{reliabilitybench2026}. AgentNoiseBench's five tool-side
categories are execution failure, incomplete response, erroneous output,
misleading signal and redundant information. None is the silent drop: a
well-formed, non-empty, plausible response that violates a constraint the caller
stated. The most dangerous mode we observe is absent from the field's model of
what can go wrong.

\paragraph{Metamorphic testing of web APIs is this work's direct ancestor}
Segura et al.\ define six Metamorphic Relation Output Patterns for RESTful web
APIs as set relations between outputs: equivalence, equality, subset, disjoint,
complete and difference~\cite{segura2018metamorphic}. Our removal differential,
in which a filter yielding the same result as no filter has been discarded, is an
instance of their equality and subset patterns, and we claim no novelty for it.
Four things distinguish this work. Their relations are identified by hand per API,
20 for Spotify and 40 for YouTube, where ours are derived mechanically from the
published schema, which is what lets the probe run unattended over hundreds of
endpoints. Their inputs are random or manually chosen, where ours model what a
language model plausibly emits. Their findings address API owners as defect
reports, where ours describe an environment property that agent builders must
design around. And they do not ask \emph{which} parameters violate relations,
which is precisely our independent variable. Broader REST fuzzing shares the first
three distinctions~\cite{atlidakis2019restler}.

\paragraph{Specification quality has been studied statically}
Recent work detects documentation and REST smells across 600 production endpoints
and states explicitly that it ``does not aim to establish causal inference
between documentation quality and agent performance''~\cite{openapismells2026}. A
study of wrapping REST APIs as MCP servers executes 2{,}190 operations across 80
specifications, but only with valid input values, and its authors observe that
their runtime failure categories ``require dynamic request-response analysis to
detect reliably''~\cite{resttomcp2025}. That analysis is what this paper
performs. Closest in mechanism is work using LLM-assisted request mutation to
generate documentation and find defects~\cite{decrop2025restnow}; its target is
the API and its documentation, whereas ours is the consumer and what reaches the
user.

\paragraph{Practitioners named the behaviour without measuring it}
API design communities distinguish \emph{lenient} handling, in which unknown or
invalid parameters are ignored, from \emph{strict} handling, in which they are
rejected. The trade-off is discussed in public API guidelines and specification
issue trackers, and practitioners note that a client typo is more likely than a
server bug. We are not aware of a prevalence study, nor of any measurement of
what lenient handling does to an automated consumer.

\section{Methodology}
\label{sec:method}

\begin{figure*}[t]
\centering
\includegraphics[width=\textwidth]{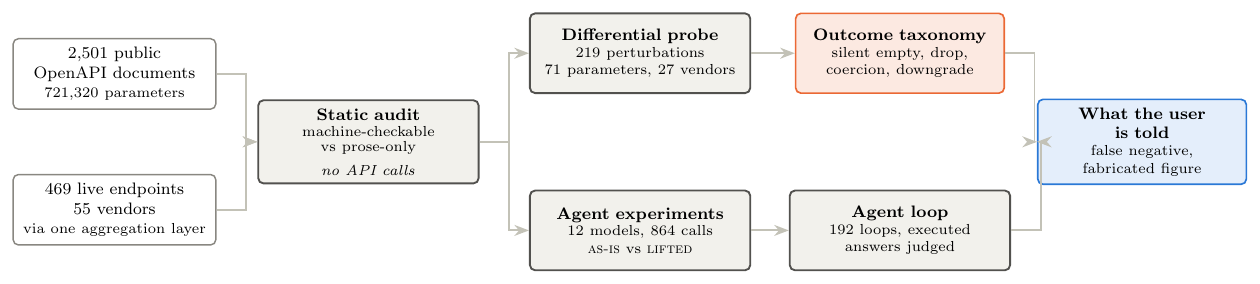}
\Description{Flow diagram of the study. Two data sources, a public OpenAPI corpus and live endpoints reached through an aggregation layer, feed a static audit, which feeds a differential probe and a set of agent experiments; the agent experiments feed an agent loop, and both branches end at an outcome taxonomy and a measure of what the user is told.}
\caption{Study overview. The static audit reads published schemas and issues no
         API calls, so it covers both corpora; the two execution stages run against
         live commercial endpoints, reached through the Monid aggregation layer,
         which returns a citable run identifier for each of them. Counts are the
         scored totals reported in Section~\ref{sec:results}.}
\label{fig:pipeline}
\end{figure*}

Three instruments, applied in order: a static audit that costs nothing, a
differential probe against live endpoints, and an agent loop that executes what a
model produces. Results for all three are in Section~\ref{sec:results}.

\subsection{Auditing constraint form}
\label{sec:auditmethod}

For each endpoint we walk the published JSON Schema and enumerate every scalar
leaf parameter. A leaf is \emph{machine-checkable} if it declares \texttt{enum},
\texttt{pattern}, \texttt{format} or numeric bounds. A leaf carries a
\emph{documentation-constraint gap} if a conservative pattern matcher finds a
controlled vocabulary, numeric bound or format stated in the description with no
corresponding schema keyword. Both classifications are deterministic functions of
published artefacts; the audit issues no API calls.

One detail matters for reproduction. The human description of an array-valued
parameter is written at the array level rather than on its items, so a walker that
does not carry descriptions down misses array-valued vocabularies entirely, which
is where most of the interesting cases are.

\subsection{Instruments}
\label{sec:instruments}

The study uses two aggregation layers, one on each axis it needs to vary. An
aggregation layer sits between a caller and many independent providers, exposing
them through a single contract, a single credential and a uniform request format,
and billing them centrally. Neither is the object of study; both are apparatus,
and each removes a barrier that would otherwise make the comparison impractical.

\textbf{Monid}~\cite{monidapi} aggregates \emph{data and tool endpoints}: web
search, scraping, people and company enrichment, social and commercial data, and
similar services from many independent vendors, reached with one key and priced
per call. It publishes a JSON Schema for each endpoint and returns a structured
record for each execution. This is the layer the tested endpoints sit behind.

\textbf{OpenRouter}~\cite{openrouterapi} aggregates \emph{language models}:
models from many providers behind one key and one OpenAI-compatible request
format, billed centrally. This is how the twelve models in
Section~\ref{sec:agentsetup} are reached.

The symmetry is the reason the design is feasible. Comparing failure behaviour
across vendors requires that vendors be callable uniformly; comparing model
behaviour across families requires that models be callable uniformly. Without
the first, the perturbation study is a procurement exercise; without the second,
the agent study is twelve separate integrations. Each layer also introduces a
dependency, and we treat both as limitations in
Section~\ref{sec:discussion}: the first constrains which endpoints are in the
sample, and the second means models are reached through a gateway rather than
each provider's native API.

\paragraph{What the method depends on}
Four properties of the endpoint layer are load-bearing, and a replication needs
an equivalent for each:

\begin{itemize}
\item \emph{Schemas in one format, retrievable without charge or
      authentication.} This is what makes the audit free and what lets the same
      classifier run over every vendor rather than one parser per vendor.
\item \emph{A run identifier, price and timing returned by every execution.}
      Each of the 219 perturbations in Section~\ref{sec:res-measure} can be
      re-fetched individually, so a reader can check any single measurement
      rather than trusting an aggregate.
\item \emph{One contract and one key across 27 vendors.} Executing thousands of
      deliberately malformed calls against commercial APIs is otherwise a
      procurement exercise before it is a research one.
\item \emph{A single schema validator in front of every vendor.} This is the
      subtle one. Because all requests pass one validator, gateway rejections and
      vendor behaviour are separable, which is what turns the intermediary from a
      confound into the mechanism (Section~\ref{sec:res-measure}).
\end{itemize}

\noindent
We run the identical audit over two populations: the endpoints reachable through
that layer, and a public corpus. The second is APIs.guru~\cite{apisguru}, a public
directory of independently published OpenAPI documents, included specifically so
that the prevalence claim does not rest on one organisation's schema-writing
habits. OpenAPI 2.0 and 3.x place a parameter's type differently and both use
\texttt{\$ref}; we resolve local references and count unresolvable ones rather
than guessing.

\subsection{The differential probe}
\label{sec:probe}

\begin{figure*}[t]
\centering
\includegraphics[width=\textwidth]{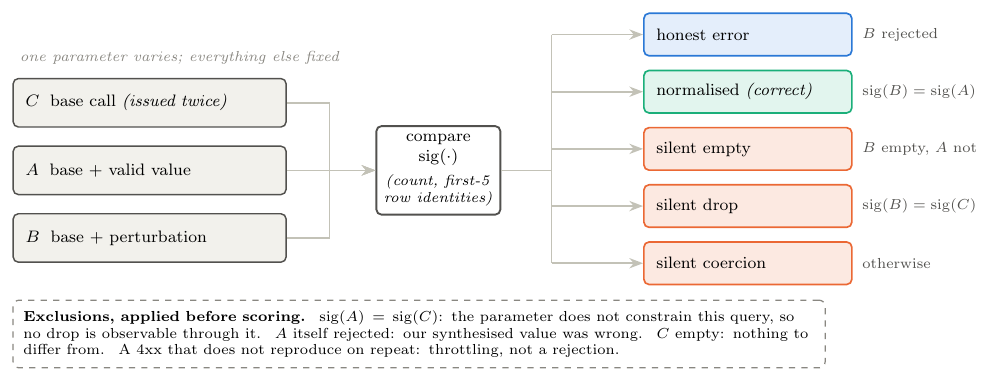}
\Description{Diagram of the differential probe. Three calls on the left, a baseline, a baseline plus a valid value and a baseline plus a perturbation, feed a signature comparison in the middle, which branches to five verdicts on the right: honest error, normalised, silent empty, silent drop and silent coercion, each labelled with its triggering condition.}
\caption{The differential probe. Three calls differ in one parameter and nothing
         else; the verdict follows from comparing their signatures. A single
         response cannot be judged, but a triple can: the equality
         $\mathrm{sig}(B)=\mathrm{sig}(C)$ is what identifies a filter the server
         accepted and discarded.}
\label{fig:probe}
\end{figure*}

For each parameter $p$ of an endpoint we issue three calls
(Figure~\ref{fig:probe}):

\begin{center}
\begin{tabular}{ll}
$C$       & the base call, shared across all $p$, issued twice \\
$A_p$     & the base call plus a valid value for $p$ \\
$B_{p,i}$ & the base call plus perturbation $i$ of that value \\
\end{tabular}
\end{center}

\noindent
Let $\mathrm{sig}(\cdot)$ be the pair (result count, fingerprint of the first five
row identities). We classify $B$ as follows. Rejected, by the gateway's schema
validator or by the vendor, is an \emph{honest error}. If
$\mathrm{sig}(B)=\mathrm{sig}(A)$ the server interpreted the variant correctly and
we record \emph{normalised}, which is correct behaviour and not a failure. If $B$
is empty while $A$ is not, \emph{silent empty}. If
$\mathrm{sig}(B)=\mathrm{sig}(C)$ the parameter contributed nothing,
\emph{silent drop}. Otherwise the value was reinterpreted, \emph{silent
coercion}.

We call this procedure, together with the perturbation families below and the
controls that follow, \emph{SilentProbe}. The name covers the automation and
the input distribution; the underlying differential relation is prior
art~\cite{segura2018metamorphic} and we claim no novelty for it.

Perturbations are derived mechanically from the schema together with one small
published synonym lexicon, in ten families: enum synonym, case, whitespace,
separator, inflection, out-of-vocabulary term, format variant, numeric
out-of-range, type mismatch, and misspelling of the parameter \emph{name}. The
families model what a language model plausibly emits rather than maximising fault
detection, which is the point of departure from random-input fuzzing.

\paragraph{Controls}
Four controls were forced by the data rather than planned, and we report them as
findings about method.

\emph{Parameter-inert.} If $\mathrm{sig}(A)=\mathrm{sig}(C)$ the parameter does
not constrain the result on this query and no drop is observable through it. Such
parameters are excluded rather than scored as passes.

\emph{Noise.} The base call is issued twice and equality is tested with a
per-endpoint tolerance calibrated from that repeat. The tolerance is not
decorative: one query returned 3{,}216{,}043 and then 3{,}216{,}042 two calls
apart.

\emph{Valid-value.} If $A_p$ is itself rejected, our synthesised value was wrong
and the parameter is dropped rather than scored.

\emph{Non-empty base.} Endpoints whose base call returns nothing are excluded,
since a differential needs something to differ from.

\paragraph{Distinguishing throttling from rejection}
One vendor throttles by returning a bare \texttt{HTTP 400} rather than
\texttt{429}, so the status code cannot separate throttling from a genuine
parameter rejection. Scoring an unconfirmed 4xx would convert our own request rate
into a finding about the endpoint. We therefore pace requests and confirm every
4xx by repeating it, since a real rejection is deterministic and throttling is
not.

\subsection{Agent experiments}
\label{sec:agentsetup}

We select parameters whose accepted vocabulary is enforced but not fully
published, recovering each vocabulary by brute force. Every recovered vocabulary
is a lower bound, since we learn only about terms we thought to try. Tasks are
written in ordinary English and deliberately avoid vendor terminology, which is
the situation an agent is actually in. Twelve models across eight families are
reached through OpenRouter (Section~\ref{sec:instruments}). Each model's emitted call
is executed against the live endpoint and classified by the probe of
Section~\ref{sec:probe}. Nothing is simulated.

Two conditions differ \emph{only} in the schema. \textsc{As-Is} presents the
published schema verbatim as a function-calling tool. \textsc{Lifted} is
identical except that the controlled vocabulary is promoted into an
\texttt{enum}.

\paragraph{Why the tool exposes three parameters}
With an endpoint's full twenty-parameter schema exposed, some models fill fields
they cannot possibly know, including the pagination cursor and a parameter this
endpoint rejects for every value. The call then fails and the parameter under
test becomes unobservable. The rate is extreme and strongly model-dependent: over
751 calls, two models filled an unguessable parameter in 100\% of their calls (58
of 58 and 67 of 67) while five others did so in 0 of 72. One filled every visible
parameter with \texttt{"*"}, a wildcard syntax it invented and the schema does not
document. Mean parameters per call was 5.4 against a task needing 3, with a
maximum of 17. Restricting the tool to the three relevant parameters brings the
mean to exactly 3.0 and removes the confound. This is a control, but it is also a
result: a model that fills every field it can see maximises its own exposure to
prose-only constraints.

\begin{figure}[t]
\centering
\includegraphics[width=\linewidth]{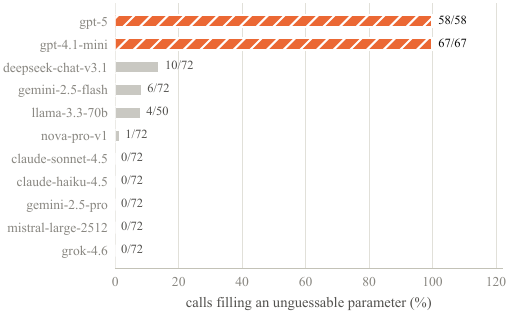}
\Description{Ranked horizontal bars showing the share of calls in which each model filled a parameter it could not know. Two models do so in 100 percent of calls, five in none.}
\caption{Share of calls in which a model filled a parameter it could not know, such as the pagination cursor, when the full twenty-parameter schema was exposed. The ordering is the finding: this is 100\% against 0\% with little in between. A model that fills every field it can see maximises its own exposure to prose-only constraints.}
\label{fig:stuffing}
\end{figure}

Models also emit parameter \emph{names} that differ from the schema property, for
instance dropping a bracket suffix. We log this drift and re-execute with names
normalised, so a wrong value and a wrong name are never confounded.

\paragraph{Downstream loop}
To measure harm rather than exposure we run the full loop: the model's call is
executed, the real response is fed back, and if it calls again that call is
executed too, up to four turns. We keep the final natural-language answer and
classify it with a judge model on whether it asserts absence, hedges, reports
zero as the answer, invents a figure, or declines. Judge labels are stored
verbatim for audit, and the judge is a different model from those under test. The
control condition issues a query that lands inside the vocabulary and returns
real rows, giving a baseline for how confidently these models assert things when
the data is genuinely there.

\section{Results}
\label{sec:results}

\subsection{How common is the gap?}
\label{sec:res-audit}

Table~\ref{tab:corpora} reports the audit over both corpora. Three findings, of
which one does not replicate.

\begin{table}[t]
\caption{The same audit over two independent corpora. Dashes mark quantities the
         aggregation-layer pass did not separate.}
\label{tab:corpora}
\small
\begin{tabular}{lrr}
\toprule
 & Aggregation layer & Public corpus \\
\midrule
documents / endpoints        & 469      & 2{,}501 \\
providers / operations       & 55       & 79{,}539 \\
parameter leaves             & 1{,}966  & 721{,}320 \\
leaf has any description     & ---      & 70.6\% \\
\midrule
declares \texttt{enum}       & 13.1\%   & 7.5\% \\
declares \texttt{pattern}/\texttt{format} & --- & 6.6\% \\
declares numeric bounds      & ---      & 1.1\% \\
\midrule
prose-only vocabulary        & 2.4\%    & 0.3\% \\
prose-only format            & 7.1\%    & 1.6\% \\
prose-only bound             & 0.2\%    & 0.5\% \\
\midrule
documents with $\geq$1 gap   & 20.9\%   & \textbf{40.1\%} \\
\bottomrule
\end{tabular}
\end{table}

\begin{figure}[t]
\centering
\includegraphics[width=\linewidth]{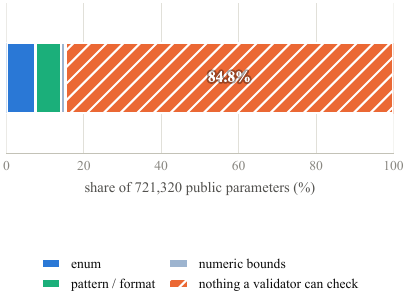}
\Description{Single stacked bar showing what governs a parameter across the public corpus: 7.5 percent declare an enum, 6.6 percent a pattern or format, 1.1 percent numeric bounds, and 84.8 percent nothing a validator can check.}
\caption{What governs a parameter, across the public corpus. The last segment is the finding: for 84.8\% of parameters there is nothing a validator can enforce, whatever the description may say. A further 29.4\% carry no description either, leaving a caller with neither a machine constraint nor a sentence to read.}
\label{fig:corpus}
\end{figure}

\textbf{Most parameters carry no machine-checkable constraint at all.} Across
721{,}320 public parameters, 7.5\% declare an enumeration and 15.2\% declare any
machine constraint whatsoever. Whatever governs the remaining 85\%, no validator
can see it. A further 29.4\% carry no description either, leaving a caller with
neither a machine constraint nor a sentence to read.

\textbf{The gap is widespread per document.} 40.1\% of public API documents state
at least one constraint in prose that their schema does not encode, higher than
the 20.9\% measured on the aggregation layer.

\textbf{The per-leaf prose-vocabulary rate does not replicate}, and the likely
reason cuts against us. At 2.4\% against 0.3\% the difference is eightfold. Many
public specifications are generated from source code, and a generator emits
\texttt{enum} because the source enumeration is right there, whereas hand-written
agent-facing descriptions narrate the vocabulary in a sentence. The specific
failure this paper concerns may therefore be more common in hand-authored agent
interfaces than in the generated specifications that dominate public corpora. We
state this as a scope limit rather than leaving it for the reader to notice.

\subsection{Does constraint form predict honesty?}
\label{sec:res-measure}

\begin{figure}[t]
\centering
\includegraphics[width=\linewidth]{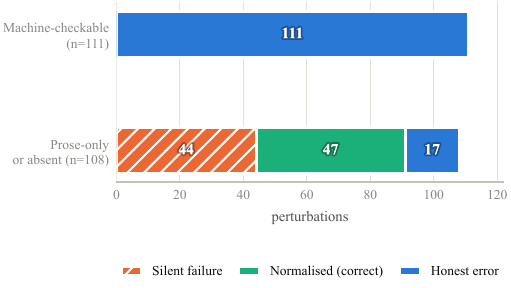}
\Description{Two stacked bars comparing outcomes. Machine-checkable constraints produce 111 honest errors and no silent failures. Prose-only constraints produce 44 silent failures, 47 correctly normalised responses and 17 honest errors.}
\caption{Outcome by constraint form. Excluded before this figure: 11
         perturbations lost to vendor throttling, 47 that returned neither a
         signature nor an error, and 37 on five endpoints whose base call did not
         reproduce on an immediate repeat. Of those 37, seventeen had been scored
         silent coercion, which on a drifting endpoint is what noise looks like;
         retaining them would have inflated the headline. Base calls reproduced
         exactly on 42 of 47 repeats.}
\label{fig:outcome}
\end{figure}

219 perturbations were scored over 71 parameters across 27 vendors.
Figure~\ref{fig:outcome} gives the outcome distribution. Excluding perturbations
the server normalised correctly, silent failure occurs in 0 of 111
machine-checkable parameters and 44 of 61 prose-only ones; Fisher's exact test
gives $p=2\times10^{-13}$.

Of 135 parameters probed, 42 were excluded as inert and 22 because their valid
value was itself rejected, leaving 71 scored.

\paragraph{Honesty has two levels}
Being rejected is not the same as being told why. All 113 gateway rejections named
the offending parameter and enumerated its legal values, for example
\texttt{expected one of "1-10"|"11-50"|...}. All 14 vendor rejections consisted of
the entire string \texttt{Invalid input: HTTP 400}. An agent can repair itself
from the first and not from the second, so self-repair after an unactionable 4xx
degenerates into guessing. Actionable messages are themselves a product of
machine-readable schemas, because only a validator that knows the legal values can
name them.

\paragraph{The aggregation layer is the mechanism, not a confound}
Reaching vendors through an intermediary looks at first like a threat to the
execution results, since the intermediary could be the thing behaving badly.
110 of the 111 honest errors on machine-checkable parameters were raised by
Monid's schema validator, and those requests never reached the vendor.
This inverts what would otherwise be a threat. A generic validator can enforce
only what the schema states: machine-readable constraints are rejected with an
actionable message, while prose-only constraints pass straight through, reach the
vendor, and fail silently there. Every silent failure we observe is therefore
vendor behaviour on a request that \emph{passed} validation, and the main claim
needs no vendor-direct control.\footnote{Gateway strictness is not uniform.
Sending an undeclared parameter to five endpoints through the same layer, one
rejected it as an unrecognised key while four ignored it and returned normal
results; one accepts and discards a parameter its schema does not declare at all.
A caller cannot rely on the layer to catch a misspelled parameter name, and cannot
tell from the published interface whether it will.}

\subsection{Do agents walk into it?}
\label{sec:res-agent}

Of 864 attempts, 815 produced a tool call and 799 were scorable. Llama-3.3
declined to call the tool in 44 of 72 attempts and is reported as abstaining
rather than scored.

\begin{figure}[t]
\centering
\includegraphics[width=\linewidth]{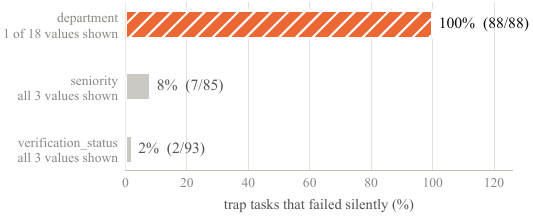}
\Description{Three horizontal bars. The parameter whose description shows 1 of 18 values fails silently on 88 of 88 attempts; the two whose descriptions state all three of their values fail on 8 and 2 percent.}
\caption{Silent-failure rate against how much of the vocabulary the description
         states. All three parameters are prose-only, so coverage is the only
         variable. Not one of twelve models ever selected a legal
         \texttt{department} value; its description reads
         \texttt{"Person department(s), e.g.\ 'executive'."}, one example out of
         eighteen accepted terms.}
\label{fig:coverage}
\end{figure}

\paragraph{Disclosure decides whether the model can comply}
The three probed parameters are all prose-only; none carries a schema constraint.
They differ only in how much of the vocabulary the description states, and that
difference dominates everything else (Figure~\ref{fig:coverage}). Partially
documented against fully documented is 88 of 88 against 9 of 178,
$p=1\times10^{-12}$.

Where the vocabulary \emph{is} written out, the same models map onto it
unprompted. Asked for people whose email is ``verified'' they emit \texttt{valid}
23 times out of 23; asked for ``catch-all'' domains they emit \texttt{accept\_all}
22 of 23; asked for vice presidents they emit \texttt{executive} 20 of 22. Models
are not bad at controlled vocabularies. They are bad at vocabularies nobody wrote
down.

\paragraph{The intervention}
Promoting the vocabulary into an \texttt{enum} takes the partially documented
parameter from 88 of 88 to 0 of 89, $p=5\times10^{-13}$. For the two
already-documented parameters there was nothing left to fix and they show the
corresponding null (2 of 93 to 0 of 90, $p=0.50$; 7 of 85 to 1 of 86, $p=0.03$).

\begin{figure*}[t]
\centering
\includegraphics[width=\textwidth]{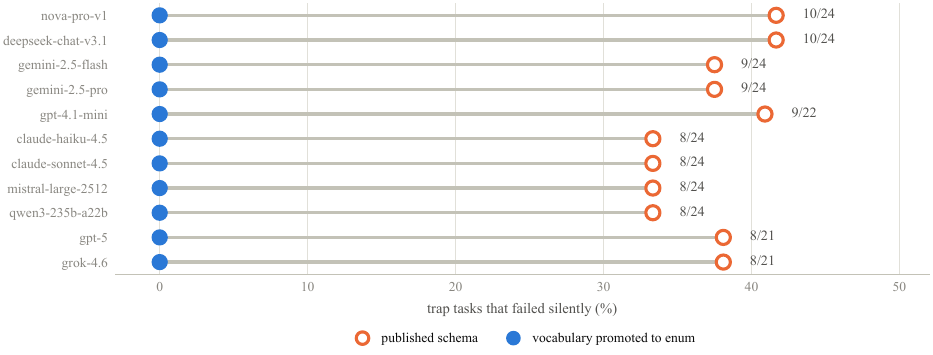}
\Description{Dumbbell chart of eleven models. Each row shows the share of trap tasks failing silently under the published schema, between 33 and 42 percent, and under the schema with the vocabulary promoted to an enum, zero for every model.}
\caption{Trap-task failure per model, published schema against the same schema
         with the vocabulary promoted to an enum. Rates are used because attempt
         counts differ slightly; raw fractions are labelled. Llama-3.3-70b is
         omitted, having declined to call the tool in 44 of 72 attempts and
         leaving 7 scorable loops.}
\label{fig:models}
\end{figure*}

Figure~\ref{fig:models} breaks the effect down by model. Eleven of twelve sit
between 8 and 10 failures out of roughly 24 under \textsc{As-Is}, and every one
falls to zero under \textsc{Lifted}. The effect does not vary by family, scale or
price. That uniformity is the point: this is a property of the interface, not of
any model's post-training.

Enumerations do constrain generation: under \textsc{Lifted}, 3 of 395 calls
emitted a value outside the declared enum. A schema enum therefore does two jobs
at once, shaping what the model emits and letting a validator reject what slips
through, and only the second survives if the model ignores it.

\subsection{What does the user get told?}
\label{sec:res-downstream}

Exposure is not harm. An agent that receives zero rows and says so cautiously has
done nothing wrong. Of 192 full loops, 94 ended on a zero-row response.

\textbf{Almost nobody notices, and nobody recovers.} 11 of 94 retried with a
different value before answering, and 0 of 11 recovered a non-zero result. The
retry traces show why:

\begin{center}\small
\begin{tabular}{ll}
\texttt{information\_security} $\to$ \texttt{security} & 0, 0 \\
\texttt{information security} $\to$ \texttt{security} $\to$ \texttt{cybersecurity} & 0, 0, 0 \\
\texttt{information security} $\to$ \texttt{security} $\to$ \texttt{infosec} & 0, 0, 0 \\
\end{tabular}
\end{center}

\noindent
The models enumerate the synonym neighbourhood, the only strategy available to
them, and the accepted term is \texttt{it}. Synonym search cannot converge on a
vocabulary it was never shown, so self-repair fails by construction rather than by
weakness.

One trace is worth the whole subsection. A model tried \texttt{engineering}, then
\texttt{engineering} again, then \emph{dropped the parameter entirely}, receiving
42{,}619{,}384 unfiltered rows, then tried \texttt{technology}. Its own repair
strategy manufactured a silent drop.

\begin{figure}[t]
\centering
\includegraphics[width=\linewidth]{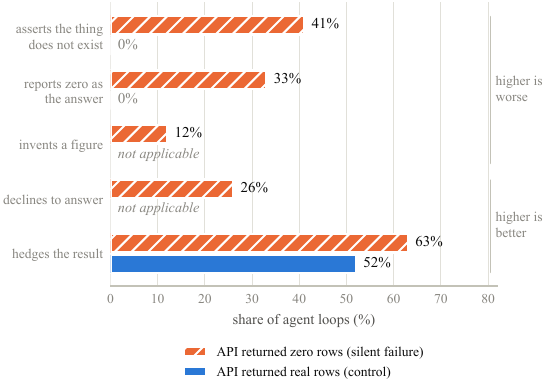}
\Description{Grouped horizontal bars comparing what agents tell users after a silent failure against a control where the API returned real rows. Asserting absence occurs in 41 percent of trap loops and none of the control loops.}
\caption{What reaches the user. Control loops issue a query inside the vocabulary
         and return real rows, which is what makes the trap column readable:
         these models do not assert absence when the data is real, so the 41\% is
         caused by the silent failure rather than by a general tendency to
         overclaim. 26 of 94 loops asserted absence with no hedge at all.}
\label{fig:downstream}
\end{figure}

Figure~\ref{fig:downstream} reports the outcome. Two model-level patterns matter
more than the averages. One model asserts absence in 9 of 10 trap loops and
hedges in none of them; several of its answers consist of the single character
\texttt{0}. Another invents a figure in 8 of its loops, answering ``5 to 7
million engineers work at US companies'' from parametric memory after the tool
returned nothing. That is the worst available outcome: the number did not come
from the data source, and nothing in the answer says so.

\textbf{The safety claim is narrower than we expected, and we state the narrow
version.} A majority of responses do hedge, so the naive framing, that agents
blindly report false negatives, is not supported by our data. What is supported: a
silent failure yields a false negative in 41\% of cases, an unhedged false
negative in 28\%, and a fabricated figure in 12\%, while the agent's ability to
detect and repair it is 12\% and 0\% respectively. Hedging is not a fix. It
transfers the problem to a user who has strictly less information than the agent
had.

\subsection{Where the effect stops}
\label{sec:boundary}

We replicated the agent experiment on two further vendors, 384 attempts and 355
tool calls, and \emph{neither produced a single silent-empty result}. That is not
a failed replication. It locates the boundary, and it exposed a mode our own
method cannot detect. Three regimes separate cleanly, by what the caller has to
go on.

\begin{figure}[t]
\centering
\includegraphics[width=\linewidth]{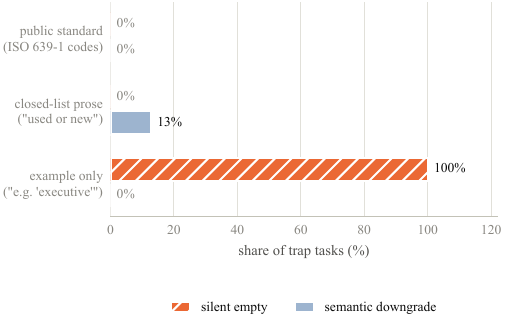}
\Description{Grouped horizontal bars over three regimes. A public standard vocabulary produces neither failure mode; a closed-list phrasing produces semantic downgrade but no silent empties; an example-only phrasing produces silent empties in every attempt.}
\caption{The two failure modes trade off across regimes. Where a closed-list phrasing prevents invention, semantic downgrade appears in its place; only the example-only phrasing produces silent empties. A public-standard vocabulary produces neither, because the model can guess it.}
\label{fig:regimes}
\end{figure}

\textbf{A public standard needs no disclosure.} One endpoint's language parameter
documents no values at all, and full language names return zero rows. It does not
matter: every model emits \texttt{en}, \texttt{es}, \texttt{fr} unprompted,
because ISO 639-1 is common knowledge. Undisclosed is dangerous only when it is
also unguessable.

\textbf{A closed-list phrasing prevents invention and causes something else.}
Another endpoint documents \texttt{"Vehicle condition: used or new."} That list is
incomplete, since \texttt{cpo} also works, but it \emph{reads} closed and models
respect it. Asked for certified pre-owned cars, 23 of 23 emitted \texttt{used}. No
zero result, no error, 21 plausible rows returned, and the wrong ones. The model
quietly weakened the request to the nearest thing the documented vocabulary could
express.

We call this a \textbf{semantic downgrade}, and it is a fourth silent mode. It is
also a blind spot of our own method: the call is well formed, non-empty, and
differs from the baseline, so the removal differential cannot see it. Detecting it
requires comparing the result against the caller's \emph{intent}, which requires
knowing the task, and is therefore invisible to any purely mechanical oracle. It
is also the mode most likely to survive into a downstream answer, because nothing
about it looks wrong. The intervention repairs it too: 23 of 179 without the enum,
0 of 176 with it.

\textbf{An example invites invention.} The third regime is the
\texttt{"e.g.\ 'executive'"} case of Section~\ref{sec:res-agent}, at 88 of 88.

\section{Discussion}
\label{sec:discussion}

\subsection{Limitations and future directions}

\textbf{Sampling.} The executed measurements cover endpoints reachable through
Monid, read-only, priced at or below \$0.02 per call, and admitting a synthesisable
valid call. This is not a random sample of production APIs. The
public corpus of Section~\ref{sec:res-audit} covers the static prevalence claim;
the execution claim does not have that cover.

\textbf{Selection bias in baseline synthesis.} Automatic value synthesis succeeds
more readily for parameters carrying an enumeration or an example, which is
correlated with the independent variable. Hand-written baselines are reported
alongside the synthesised ones, and the audit issues no calls at all.

\textbf{The inert exclusion biases downward.} $\mathrm{sig}(A)=\mathrm{sig}(C)$
arises both when a parameter does not discriminate on this query and when it is
\emph{always} discarded, including for the valid value. A single differential
cannot separate them, so the 42 parameters excluded as inert may contain genuine
silent drops. The direction of the bias is knowable: it can only make our silent
rate an underestimate.

\textbf{Detecting silent drop requires a large base result set.} Of 19 free-text
parameters probed for always-discarded behaviour, 9 appeared discarded but only 2
were conclusive, both on result sets exceeding 246 million rows. The remaining 7
sat on 2 to 50 rows, where a filter having no visible effect proves nothing. This
is a selection criterion for future work, not a detail.

\textbf{Coverage asymmetry.} The agent experiments cover three vendors against 27
for the measurement campaign. The two halves have complementary rather than equal
coverage. Model access is via one gateway rather than each vendor's native API,
which could in principle alter request handling.

\textbf{Vocabulary recovery is brute force}, so every reported vocabulary and
every count of silently failing terms is a lower bound.

\textbf{Competing interest.} Monid~\cite{monidapi}, used here as the measurement
instrument, is operated by an organisation with which the authors are affiliated. We state this plainly rather than in a footnote. Two
considerations bear on it: the findings are unflattering to that layer's upstream
vendors rather than favourable to the layer, and Section~\ref{sec:res-audit}
exists specifically so that the prevalence claim does not depend on it. No vendor
was given advance notice of, or influence over, these results.

\subsection{Recommendations}

\paragraph{For API publishers}
Put the vocabulary in the schema. If you will not, write a closed list rather than
an example: \texttt{"e.g."} is the dangerous phrasing, and a closed list keeps
callers inside the documented set even when that set is incomplete. Separately,
distinguish ``no matches'' from ``this filter is not supported here.'' No HTTP
status makes that distinction, and the two are indistinguishable to every
automated consumer.

\paragraph{For tool-layer and MCP authors}
Expose fewer parameters. Models fill what they can see, and every additional
prose-constrained field is exposure. Validate against the schema and return the
legal values in the error message: our data shows the difference between an
actionable and an unactionable rejection is the difference between repair and
guessing.

\paragraph{For agent builders}
A zero-row result on a filtered query is not evidence of absence. Re-issuing the
query without the filter distinguishes the two cases cheaply, but note that the
unfiltered retry is itself a silent drop if its result is then reported as though
filtered, which is exactly what we observed one model do.

\paragraph{For benchmark authors}
Simulated environments cannot exhibit this class of failure, because a simulator
generates responses from documentation and documentation does not describe its own
omissions. A robustness benchmark built on simulated tools reports a robustness it
has not tested.

\paragraph{On automation}
The gap is mechanically detectable from published artefacts, and automated
documentation repair is an active line of work~\cite{decrop2025restnow}. Closing
it at scale looks tractable.

\section{Conclusion}

Silent failure in agent tool calls is predicted by how a constraint is published
rather than by which vendor serves it. Two separable properties govern it.
Disclosure determines whether a model can select a legal value: a vocabulary
written out in prose is used correctly, one shown by example is not.
Machine-readability determines whether a wrong value is caught, and it is the only
form of a constraint that anything other than the model can enforce. The failure
lives where a constraint is neither disclosed nor checkable, and across 721{,}320
public parameters that combination is common. The remedy is one line of schema.

\begin{acks}
\end{acks}

\nocite{*}

\bibliographystyle{ACM-Reference-Format}
\bibliography{refs}

\appendix

\section{Reproducibility}
\label{app:repro}
Every executed call carries a Monid run identifier. All are released with the
data at \url{https://github.com/Jasper0122/silentprobe}, so any individual measurement in this paper can be re-fetched and checked
against what we report, rather than taken on trust. This is a stronger
reproducibility guarantee than a live-API study normally offers: the usual
position is that the endpoint may have changed and the original response is
gone. The complete campaign cost under US\$8: US\$4.67 of
data-API spend covering schema retrieval, endpoint verification and the scored
perturbations, plus US\$1.89 of model inference through OpenRouter.

Caching keeps the agent experiments cheap, because models converge on identical
arguments: 815 tool calls reduced to 44 distinct API calls and 88 executions
rather than 2{,}445. A replication after a fix therefore costs minutes.

\paragraph{A hazard worth documenting}
One vendor throttles by returning a bare \texttt{HTTP 400} rather than
\texttt{429}. Status alone cannot separate throttling from a genuine parameter
rejection, and scoring an unconfirmed 4xx would convert our own request rate
into a finding about the endpoint. We pace requests at four-second intervals and
confirm every 4xx by repeating it, since a real rejection is deterministic and
throttling is not. Replications that skip this will overstate honest-error rates.

\section{Perturbation Families}
\label{app:perturb}
Every perturbation is a deterministic function of the published schema plus
the synonym lexicon in Table~\ref{tab:lexicon}, so a reader can regenerate the
exact perturbation set from the schema dump we release. Table~\ref{tab:families}
lists the families with a worked example from the seed endpoint.

\begin{table}[h]
\caption{The ten perturbation families.}
\label{tab:families}
\small
\begin{tabular}{lll}
\toprule
Family & Derivation & Example \\
\midrule
enum synonym & published lexicon & \texttt{vp} $\to$ \texttt{vice\_president} \\
case & mechanical & \texttt{vp} $\to$ \texttt{VP} \\
whitespace & mechanical & \texttt{vp} $\to$ \texttt{\char32 vp} \\
separator & mechanical & \texttt{c\_suite} $\to$ \texttt{c-suite} \\
inflection & mechanical & \texttt{vp} $\to$ \texttt{vps} \\
out-of-vocabulary & fixed token & \texttt{unspecified\_other} \\
format variant & schema pattern & \texttt{51,200} $\to$ \texttt{51-200} \\
numeric range & schema bounds & \texttt{per\_page=3} $\to$ \texttt{0} \\
type mismatch & schema type & \texttt{3} $\to$ \texttt{'3'} \\
parameter name & property name & \texttt{headcount} $\to$ \texttt{headcount\_range} \\
\bottomrule
\end{tabular}
\end{table}

\begin{table}[h]
\caption{The published synonym lexicon, in full. These are the substitutions
         a language model plausibly emits having read a controlled vocabulary
         in prose rather than as a machine-checkable enumeration.}
\label{tab:lexicon}
\small
\begin{tabular}{ll}
\toprule
Schema value & Substituted with \\
\midrule
\texttt{asc} & \texttt{ascending} \\
\texttt{c\_suite} & \texttt{c-level} \\
\texttt{desc} & \texttt{descending} \\
\texttt{editorial} & \texttt{editorials} \\
\texttt{en} & \texttt{english} \\
\texttt{entity} & \texttt{entities} \\
\texttt{gb} & \texttt{uk} \\
\texttt{ja} & \texttt{japanese} \\
\texttt{newest} & \texttt{latest} \\
\texttt{news} & \texttt{newspaper} \\
\texttt{press\_release} & \texttt{pressrelease} \\
\texttt{recency} & \texttt{recent} \\
\texttt{relevance} & \texttt{relevant} \\
\texttt{social} & \texttt{social\_media} \\
\texttt{svp} & \texttt{senior\_vice\_president} \\
\texttt{us} & \texttt{usa} \\
\texttt{vp} & \texttt{vice\_president} \\
\texttt{zh} & \texttt{chinese} \\
\bottomrule
\end{tabular}
\end{table}

\section{Recovered Vocabularies}
\label{app:vocab}
Each vocabulary below was recovered by brute force and is a \emph{lower bound}:
we learn only about terms we thought to try. The right-hand column is what makes
the trap a trap, since every term in it is one a caller would reasonably use.

\begin{table}[h]
\caption{Vendor A4 \texttt{department}: accepted terms against terms that return zero rows
         with status 200.}
\small
\begin{tabular}{p{0.44\columnwidth}p{0.44\columnwidth}}
\toprule
Accepted & Silently returns zero \\
\midrule
\texttt{executive}, \texttt{hr}, \texttt{it}, \texttt{legal}, \texttt{marketing}, \texttt{operations}, \texttt{sales}, \texttt{support}, \texttt{communication}, \texttt{education}, \texttt{design}, \texttt{product}, \texttt{research}, \texttt{health}, \texttt{management}, \texttt{administrative}, \texttt{consulting} & \texttt{finance}, \texttt{human\_resources}, \texttt{communications}, \texttt{engineering}, \texttt{engineering\_technical}, \texttt{technical}, \texttt{business\_development}, \texttt{customer\_service}, \texttt{media}, \texttt{recruiting}, \texttt{security}, \texttt{data}, \texttt{science} \\
\bottomrule
\end{tabular}
\end{table}

\begin{table}[h]
\caption{Vendor A4 \texttt{verification\_status}: accepted terms against terms that return zero rows
         with status 200.}
\small
\begin{tabular}{p{0.44\columnwidth}p{0.44\columnwidth}}
\toprule
Accepted & Silently returns zero \\
\midrule
\texttt{valid}, \texttt{accept\_all}, \texttt{unknown} & \texttt{verified}, \texttt{VALID}, \texttt{confirmed}, \texttt{deliverable}, \texttt{risky}, \texttt{invalid}, \texttt{catch\_all}, \texttt{accept-all}, \texttt{valid\_email} \\
\bottomrule
\end{tabular}
\end{table}

\begin{table}[h]
\caption{Vendor A4 \texttt{seniority}: accepted terms against terms that return zero rows
         with status 200.}
\small
\begin{tabular}{p{0.44\columnwidth}p{0.44\columnwidth}}
\toprule
Accepted & Silently returns zero \\
\midrule
\texttt{junior}, \texttt{senior}, \texttt{executive} & \texttt{mid}, \texttt{mid\_level}, \texttt{entry}, \texttt{entry\_level}, \texttt{lead}, \texttt{manager}, \texttt{director}, \texttt{vp}, \texttt{c\_level}, \texttt{intern}, \texttt{principal}, \texttt{staff} \\
\bottomrule
\end{tabular}
\end{table}

\begin{table}[h]
\caption{Vendor V \texttt{condition}: accepted terms against terms that return zero rows
         with status 200.}
\small
\begin{tabular}{p{0.44\columnwidth}p{0.44\columnwidth}}
\toprule
Accepted & Silently returns zero \\
\midrule
\texttt{new}, \texttt{used}, \texttt{cpo} & \texttt{certified}, \texttt{certified\_pre\_owned}, \texttt{refurbished}, \texttt{pre\_owned}, \texttt{preowned}, \texttt{second\_hand}, \texttt{demo}, \texttt{any} \\
\bottomrule
\end{tabular}
\end{table}

\begin{table}[h]
\caption{Vendor A5 \texttt{job\_type}: accepted terms against terms that return zero rows
         with status 200.}
\small
\begin{tabular}{p{0.44\columnwidth}p{0.44\columnwidth}}
\toprule
Accepted & Silently returns zero \\
\midrule
\texttt{full\_time}, \texttt{part\_time}, \texttt{contract}, \texttt{temporary}, \texttt{volunteer}, \texttt{fellowship}, \texttt{permanent}, \texttt{job}, \texttt{volop} & \texttt{internship}, \texttt{freelance}, \texttt{internships} \\
\bottomrule
\end{tabular}
\end{table}

\begin{table}[h]
\caption{Vendor P \texttt{condition}: accepted terms against terms that return zero rows
         with status 200.}
\small
\begin{tabular}{p{0.44\columnwidth}p{0.44\columnwidth}}
\toprule
Accepted & Silently returns zero \\
\midrule
\texttt{new}, \texttt{used}, \texttt{certified}, \texttt{cpo}, \texttt{any} & --- \\
\bottomrule
\end{tabular}
\end{table}

\section{Per-Vendor Outcomes}
\label{app:vendors}
Vendors are anonymised. The mapping is released with the data. A vendor
contributing only machine-checkable parameters cannot exhibit a silent failure
through them, which is the point of the table rather than an omission.

\begin{table}[h]
\caption{Perturbation outcomes by vendor and constraint form.}
\label{tab:vendors}
\small
\begin{tabular}{lrrrrr}
\toprule
 & \multicolumn{2}{c}{prose-only} & \multicolumn{3}{c}{machine-checkable} \\
\cmidrule(lr){2-3}\cmidrule(lr){4-6}
Vendor & silent & honest & silent & honest & normalised \\
\midrule
Vendor A4 & 7 & 12 & 0 & 35 & 13 \\
Vendor I & 12 & 0 & 0 & 12 & 5 \\
Vendor A20 & 0 & 1 & 0 & 18 & 4 \\
Vendor A14 & 0 & 0 & 0 & 22 & 0 \\
Vendor T & 2 & 2 & 0 & 15 & 0 \\
Vendor P & 12 & 0 & 0 & 0 & 6 \\
Vendor A23 & 0 & 0 & 0 & 10 & 4 \\
Vendor V & 11 & 0 & 0 & 0 & 0 \\
Vendor X & 0 & 0 & 0 & 10 & 0 \\
Vendor A22 & 6 & 0 & 0 & 0 & 1 \\
Vendor M & 2 & 0 & 0 & 0 & 4 \\
Vendor L & 0 & 3 & 0 & 0 & 0 \\
Vendor A16 & 2 & 0 & 0 & 0 & 1 \\
Vendor K & 2 & 0 & 0 & 0 & 1 \\
Vendor S & 2 & 0 & 0 & 0 & 1 \\
Vendor A5 & 2 & 0 & 0 & 0 & 1 \\
Vendor A7 & 0 & 0 & 0 & 0 & 3 \\
Vendor A29 & 2 & 0 & 0 & 0 & 1 \\
Vendor A13 & 1 & 0 & 0 & 0 & 1 \\
Vendor A & 0 & 0 & 0 & 0 & 2 \\
Vendor A26 & 0 & 0 & 0 & 0 & 2 \\
Vendor H & 1 & 0 & 0 & 0 & 0 \\
\bottomrule
\end{tabular}
\end{table}

\section{Prompt Templates}
\label{app:prompts}
The agent experiments use one system prompt and one task string per case. No
few-shot examples, no chain-of-thought instruction and no mention of the
vendor's terminology: the point is what a model does with the interface as
published.

\paragraph{System prompt, tool-call experiments}
\begin{quote}\small\ttfamily
You are an agent with one tool. Call it to answer the user. Use only the
tool's documented parameters.
\end{quote}

\paragraph{System prompt, full agent loop}
\begin{quote}\small\ttfamily
You are a research assistant with one tool. Use it to answer the user's
question, then give the user a short direct answer.
\end{quote}

\paragraph{Task strings}
Each is a complete user turn. The trap tasks name a concept in ordinary
English whose obvious rendering is not in the vendor's vocabulary; the control
tasks name one whose rendering is.

\begin{quote}\small\ttfamily
Find engineers at US companies. Return at most 3 rows.\\[2pt]
Find people working in human resources at US companies. Return at most 3 rows.\\[2pt]
Find customer service staff at US companies. Return at most 3 rows.\\[2pt]
Find information security staff at US companies. Return at most 3 rows.\\[2pt]
Find marketing people at US companies. Return at most 3 rows.\ \textrm{(control)}\\[2pt]
Find certified pre-owned cars for sale near 90210. Return at most 3 rows.\\[2pt]
List the available English voices. Return at most 3 rows.\\[2pt]
\end{quote}

\paragraph{Judge prompt, downstream experiment}
The judge is a different model from any under test, and sees the task and the
answer but not the tool call, so it cannot infer the verdict from the query.

\begin{quote}\small\ttfamily
A research assistant was asked a question, used a search API, and gave this
answer. The API returned ZERO matching rows with a success status. The zero
was caused by a vocabulary mismatch in the query, not by an absence of such
people in the world.
\\[4pt]
QUESTION: \{task\}
\\[4pt]
ASSISTANT'S ANSWER: \{answer\}
\\[4pt]
Classify the answer. Reply with ONLY a JSON object: asserts\_absence,
hedged, reports\_zero\_as\_fact, fabricated\_number, refused\_to\_answer,
one\_line.
\end{quote}

\section{Agent Answers, Verbatim}
\label{app:answers}
Every answer below follows a call that returned HTTP 200 with zero rows and no
error field, on a question whose true answer is in the millions. Models are
identified by family only. These are transcribed exactly, including length.

\paragraph{Asserted absence with no hedge}
\begin{quote}\small
\textit{google, asked ``How many people work in engineering roles at US companies?''}\\[3pt]
``Based on the available information, there are no people working in engineering roles at US companies.''
\end{quote}
\begin{quote}\small
\textit{google, asked ``How many people work in engineering roles at US companies?''}\\[3pt]
``There are no people working in engineering roles at US companies.''
\end{quote}
\begin{quote}\small
\textit{google, asked ``How many people work in human resources at US companies?''}\\[3pt]
``There are no people who work in human resources at US companies.''
\end{quote}

\paragraph{Invented a figure the tool did not return}
\begin{quote}\small
\textit{meta-llama, asked ``How many people work in human resources at US companies?''}\\[3pt]
``About 120,000 people work in human resources at US companies.''
\end{quote}
\begin{quote}\small
\textit{mistralai, asked ``How many people work in engineering roles at US companies?''}\\[3pt]
``**Approximately 5–7 million people** work in engineering roles at US companies. For precise real-time data, a specialized labor database or government source like the U.S. Bureau of Labor Statistics would be needed.''
\end{quote}

\paragraph{Hedged, which is the majority behaviour}
\begin{quote}\small
\textit{anthropic, asked ``How many people work in engineering roles at US companies?''}\\[3pt]
``The search returned 0 results for people in engineering roles at US companies. This could mean: 1. There are no matching records in the database with these Vendor Xct filters 2. The system may need a more specific company filter to return results 3. The database may not have comprehensive coverage of all US companies **Answer: 0 people found** with the current search parameters in this database. Note: This doesn't mean there are zero engineers at US [\ldots]''
\end{quote}
\begin{quote}\small
\textit{anthropic, asked ``How many people work in engineering roles at US companies?''}\\[3pt]
``The search returned 0 results for engineering roles at US companies. This could mean: 1. There are no matching records in the database with those specific filters 2. The database may not have broad coverage of all US companies 3. The department filter value might need to be different **Answer: 0** - Based on the available data in this system, there are no engineering roles at US companies currently indexed. This likely reflects limitations in the [\ldots]''
\end{quote}

\section{Retry Traces}
\label{app:retries}
Every retry in the downstream experiment, with the value tried at each step
and the row count returned. None recovered. The accepted term for the security
cases is \texttt{it}, which no synonym search reaches.

\begin{table}[h]
\small
\begin{tabular}{llp{0.40\columnwidth}l}
\toprule
Family & Task & Values tried & Counts \\
\midrule
deepseek-chat-v3.1 & security & \texttt{information security}, \texttt{security}, \texttt{cybersecurity} & 0, 0, 0 \\
mistral-large-2512 & engineer & \texttt{engineering}, \texttt{engineering} & 0, 0 \\
grok-4.6 & security & \texttt{information security}, \texttt{security} & 0, 0 \\
nova-pro-v1 & security & \texttt{information security}, \texttt{security} & 0, 0 \\
nova-pro-v1 & security & \texttt{information security}, \texttt{security}, \texttt{infosec} & 0, 0, 0 \\
mistral-large-2512 & verified & \texttt{valid}, \texttt{valid} & ---, --- \\
grok-4.6 & engineer & \texttt{engineering}, \texttt{engineering} & ---, 0 \\
\bottomrule
\end{tabular}
\caption{Retry sequences. The third row is the trace in which a model dropped
         the filter entirely and received 42{,}619{,}384 unfiltered rows.}
\label{tab:retries}
\end{table}

\end{document}